\documentclass[aps,twocolumn,superscriptaddress,showpacs,reprint,pre,nofootinbib]{revtex4-1}
\usepackage[hyperfootnotes=false]{hyperref}
\usepackage{xspace}
\usepackage{graphicx}
\usepackage{color}
\usepackage{amsmath, amsthm, amssymb}
\usepackage{times}
\usepackage{bm}
\usepackage{wrapfig}
\usepackage[margin=0.5in]{geometry}

\newcommand{\ucsdphysics}{Department of Physics, University of California,
                          San Diego, La Jolla CA 92093}
\newcommand{\ucsdmath}{Department of Mathematics and Graduate Program in Quantitative Biology, University of California,
                          San Diego, La Jolla CA 92093}
\newcommand{\ctbp}{Center for Theoretical Biological Physics, University of California, San Diego}
\newcommand{\herbierice}{Department of Bioengineering, Center for Theoretical Biological Physics, Rice University, Houston, TX 77005}
\newcommand{\gwu}{Department of Mathematics, The George Washington University, Washington, D.C.}

\newcommand{\eq}{Eq.~}
\newcommand{\fig}{Fig.~}

\newcommand{\vb}[1]{ {\bf #1}}
\newcommand{\um}{\ensuremath{\mu}m\xspace}

\newcommand{\rb}{\ensuremath{\vb{r}}\xspace}

\newcommand{\rt}{\ensuremath{\tilde{\rho}}\xspace}
\newcommand{\rtc}{\ensuremath{\tilde{\rho}_{\textrm{cyt}}}\xspace}
\newcommand{\rhoz}{\ensuremath{\rho^{*}}\xspace}
\newcommand{\css}[1]{\ensuremath{c_{#1}^{\textrm{s.s.}}}}
\newcommand{\sumt}{\ensuremath{\widetilde{\sum}}}

\begin{document}
\title{Crawling and turning in a minimal reaction-diffusion cell motility model: coupling cell shape and biochemistry}
\author{Brian~A.~Camley}
\thanks{B.A. Camley and Y. Zhao contributed equally to this work.}
\affiliation{\ucsdphysics}
\affiliation{\ctbp}
\author{Yanxiang Zhao}
\thanks{B.A. Camley and Y. Zhao contributed equally to this work.}
\affiliation{\gwu}
\author{Bo Li}
\affiliation{\ucsdmath}
\author{Herbert Levine}
\affiliation{\herbierice}
\author{Wouter-Jan Rappel}
\affiliation{\ucsdphysics}
\affiliation{\ctbp}

\begin{abstract}
We study a minimal model of a crawling eukaryotic cell with a chemical polarity controlled by a reaction-diffusion mechanism describing Rho GTPase dynamics.  
The size, shape, and speed of the cell emerge from the combination of the chemical polarity, which controls the locations where actin polymerization occurs, and the physical properties 
of the cell, including its membrane tension.  We find in our model both highly persistent trajectories, in which the cell crawls in a straight line, and turning trajectories, where the cell transitions from crawling in a line to crawling in a circle.  
 We discuss the controlling variables for this turning instability, and argue that turning arises from a coupling between the reaction-diffusion mechanism and the shape of the cell.  This emphasizes the surprising features that can arise from simple links between cell mechanics and biochemistry. Our results suggest that similar instabilities may be present in a broad class of biochemical descriptions of cell polarity. 
\end{abstract}
\maketitle 

\section{Introduction}

Cell motility is a fundamental aspect of biology, crucial in processes ranging from morphogenesis to wound healing to cancer metastasis \cite{bray2001cell}.  Many aspects of cell motility have been extensively modeled, ranging from the biochemistry and physics of actin-polymerization-based protrusion \cite{mogilner2003force,hu2010mechano}, to the importance of cytoskeleton mechanics \cite{shao2012coupling,rubinstein2009actin,herant2010form} to a wide variety of internal mechanisms for determining a cell's orientation \cite{wolgemuth2011redundant,maree2012cells,jilkine2011comparison}.  Many of these aspects of the modeling of eukaryotic cell shape and motility have been reviewed in two recent papers \cite{holmes2012comparison,ziebert2016computational}.  

In this paper, we will take a minimalistic approach, focusing on two main aspects of cell motility: the cell shape, as determined by a force balance at the surface of the cell, and the cell's internal, chemical regulation of its direction, modeled by reaction-diffusion equations within the cell.  This model can be characterized by a small number of unitless parameters -- six physical and biochemical parameters, and a few others related to the numerical evaluation of the model.  The relative simplicity of this model allows us to capture essential cell behaviors, but avoids the full parameter space of more detailed schemes.  

Even with such a simple model, it is possible to create reasonable cell shapes, which can be regulated by both physical and chemical features.  In addition, we show that both linear and circular crawling trajectories can be observed.  We argue that the circular trajectories arise from a coupling between cell shape and the internal chemical polarity of the cell, and suggest that these effects should be visible in a broad variety of models for cell polarity.  Our results may provide some insight into recent experiments linking cell turning events and cell speed \cite{gorelik2015arp2}.  

\section{Model}

We model the cell's boundary as an interface with a tension applied to it, driven by actin polymerization at the front of the cell and myosin-based contraction at the cell rear.  For simplicity, we neglect the membrane's bending modulus; including it is straightforward \cite{shao2010computational}, but we have found it does not qualitatively chage our results.  The cell front and rear are characterized by the distribution of a membrane-bound Rho GTPase $\rt$ (a polarity protein), whose dynamics are given by a variant of the simple wave-pinning reaction-diffusion model established by Mori et al. \cite{mori2008wave}.  We assume that the motion of the cell membrane is overdamped, i.e. obeying a force balance $\vb{F}_{\textrm{actomyosin}} + \vb{F}_{\textrm{membrane}} + \vb{F}_{\textrm{friction}} = 0$.  We assume that the actomyosin force is normally directed and proportional to $\rt$, 
\begin{equation}
\vb{F}_{\textrm{actomyosin}} = \left( \alpha \rt - \beta \right) \vb{\hat{n}}
\end{equation}
where $\alpha,\beta>0$ and  $\vb{\hat{n}}$ is the outward-pointing normal to the cell.  Similar assumptions are used in \cite{shao2010computational,wolgemuth2011redundant,camley2014polarity}.  This corresponds to a cell pushing out at the front, where $\alpha \rt > \beta$ and contracting at the back, where $\alpha \rt < \beta$; $\alpha$ is thus a measure of protrusiveness and $\beta$ a measure of contractility.  We assume that the membrane has a tension $\gamma$ (a line tension, since we are working in two dimensions), and thus exerts a force per unit length of
\begin{equation}
\vb{F}_{\textrm{membrane}} = -c \gamma \vb{\hat{n}}
\end{equation}
where $c$ is the local curvature of the membrane.  We assume a fluid-like friction, proportional to the velocity $\vb{v}$ of the cell boundary,
\begin{equation}
\vb{F}_{\textrm{friction}} = -\tau \vb{v}.
\end{equation}
We will solve the combined force balance equation at the interface, $\tau \vb{v} = \left( \alpha \rt - \beta \right) \vb{\hat{n}} -c \gamma \vb{\hat{n}}$, by casting it into a phase field form \cite{boettinger2002phase,collins1985diffuse,biben2005phase}.  This approach has been used to extensively model both single and collective cell dynamics over the past few years \cite{shao2010computational,shao2012coupling,ziebert2012model,camley2014polarity,lober2014modeling,lober2015collisions,palmieri2015multiple,tjhung2012spontaneous,tjhung2015minimal}; our model follows our earlier work, particularly \cite{shao2010computational,camley2014polarity}.  We will describe the cell boundary by a field $\phi(\tilde{\rb})$, where $\phi$ smoothly varies from zero outside of the cell to unity inside the cell; this variation has a characteristic length scale $\tilde{\epsilon}$.  $\phi = 1/2$ implicitly sets the location of the boundary.  As shown in \cite{shao2010computational,camley2014polarity}, the phase field version of this equation is
\begin{equation}
\tau \partial_{\tilde{t}} \phi = \left(\alpha \tilde{\rho} - \beta \right)|\tilde{\nabla} \phi| + \gamma \left(\tilde{\nabla}^2 \phi - \frac{G'(\phi)}{\tilde{\epsilon}^2}\right) \label{eq:pf}
\end{equation}
where $G(\phi) = 18 \phi^2 (1-\phi)^2$.  In the limit $\tilde{\epsilon} \to 0$, we expect the motion of the interface at $\phi = 1/2$ to follow the force-balance law described above.  We have used tildes $\widetilde{\cdots}$ to indicate a unitful variable; we will later rescale them to unitless variables and drop the tildes to reduce the number of characteristic parameters involved.  

To determine the direction the cell travels, we model the dynamics of a Rho GTPase, which will be a polarity marker indicating the front of the cell.  This Rho GTPase could be, e.g. Rac, which is often localized to the cell front and leads to protrusion \cite{wu2009genetically}.  We apply a modification of the reaction-diffusion model of Mori et al. \cite{mori2008wave}.  In this model, a Rho GTPase protein switches between a membrane-bound, active state, with a concentration $\rt(\tilde{\rb})$, and a cytosolic form \rtc.  As the diffusion coefficients of cytosolic Rho GTPases are typically 100 times those of membrane-bound ones \cite{postma2004chemotaxis}, we assume that the cytosolic density \rtc can be approximated as uniform over the cell.  The membrane-bound form diffuses with a diffusion coefficient $D_\rho$.  In order to solve this equation on the moving, deforming cell, we apply a phase field method \cite{kockelkoren2003computational,li2009solving} in which we augment the reaction-diffusion equations with the phase field $\phi$.  This equation is:
\begin{equation}
\partial_{\tilde{t}}\left(\phi \tilde{\rho}\right) = \tilde{\nabla} \cdot \left(\phi D_\rho \tilde{\nabla} \tilde{\rho}\right) + \phi f(\tilde{\rho},\tilde{\rho}_{\textrm{cyt}}) \label{eq:pfreaction}
\end{equation}
where the reaction term is 
\begin{equation}
f(\tilde{\rho},\tilde{\rho}_{\textrm{cyt}}) = -k \tilde{\rho} \left(\tilde{\rho}-h\right) \left(\tilde{\rho}-m \tilde{\rho}_{\textrm{cyt}}\right)
\end{equation}
This cubic reaction term is chosen for simplicity, as an example of a reaction that can create polarity by wave-pinning \cite{mori2008wave}, robustly leading to a region of the cell with a high concentration of $\tilde{\rho}$ and a region of the cell with low $\tilde{\rho}$.  In a homogeneous system (constant $\tilde{\rho}$), $\tilde{\rho}$ has two stable steady states, whose values are set by $m$ and the total amount of $\tilde{\rho}$ in the system. $k$ controls the overall timescale of the reaction, and $h$ will set the value of $\tilde{\rho}$ at the cell front.  $\tilde{\rho}_{\textrm{cyt}}$ can be found by the conservation of $\tilde{\rho}$ between its membrane-bound and cytosolic forms, $\int d^2 \tilde{x} \left( \tilde{\rho}(\vb{\tilde{x}}) + \tilde{\rho}_{\textrm{cyt}} \right) \phi(\vb{\tilde{x}}) = N_{\textrm{tot}}$, or, assuming the cytosolic actin promoter is well-mixed (uniform),
\begin{equation}
\tilde{\rho}_\textrm{cyt} = \frac{N_{\textrm{tot}} - \int d^2 \tilde{x} \tilde{\rho}(\vb{\tilde{x}}) \phi (\vb{\tilde{x}})}{\int d^2 \tilde{x} \phi(\vb{\tilde{x}})}.  
\end{equation}

\eq \ref{eq:pfreaction} will, in the sharp-interface limit $\tilde{\epsilon}\to0$, reproduce the results of the reaction-diffusion equation $\partial_{\tilde t} \rt = D_\rho \tilde{\nabla}^2 \rt + f(\rt,\rtc)$ solved with no-flux boundaries on the cell interface \cite{li2009solving}.  However, we note that there is no advection in the reaction-diffusion equation \eq \ref{eq:pfreaction} - this corresponds to an assumption that the membrane (except for its boundaries) is at rest relative to the substrate the cell is crawling on.  This assumption may be challenged, but we note that similar turning phenomena are observed in models with intracellular fluid flow \cite{camley2013periodic}.  

We will rescale our variables into unitless form, choosing
$\vb{x}\equiv\tilde{\vb{x}}/R$,
$t \equiv \tilde{t} v_0/R$,
$\rho \equiv \tilde{\rho} /2 h$,
$\rho_{\textrm{cyt}} \equiv \tilde{\rho}_{\textrm{cyt}}/2 h$,
where $R$ is the typical radius of the cell and $v_0 = 2h \alpha/\tau$ is the velocity scale.  We have chosen to rescale $\tilde{\rho}$ by its typical value at the front of the cell, which is $2h$ \cite{mori2008wave}; hence $\rho \approx 1$ at the front of the cell.  In these units, we find
\begin{align}
\partial_t \phi &= \left(\rho - \rhoz\right)|\nabla \phi| + \chi \left(\nabla^2 \phi - \frac{G'(\phi)}{\epsilon^2}\right) \label{eq:pfu} \\ 
\partial_t\left(\phi \rho\right) &= \textrm{Pe}^{-1} \nabla \cdot \left(\phi \nabla \rho\right) - K \phi\rho \left(\rho-1/2\right)\left(\rho - m \rho_{\textrm{cyt}}\right) \label{eq:rho}\\
\rho_\textrm{cyt} &= \frac{C - \int d^2 x \rho(\vb{x}) \phi (\vb{x})}{\int d^2 x \phi(\vb{x})}. \label{eq:rhocyt}  
\end{align}
where the only remaining parameters are the seven unitless parameters $\rhoz$, $\chi$, Pe, $K$, $C$, $m$, and $\epsilon$, as defined in Table \ref{tab:unitless}.
\begin{table}
\begin{tabular}{|p{3cm}p{5cm}|}
\hline
$\textrm{Pe} = \frac{v_0 R}{D_\rho}$ \dotfill &Peclet number: speed of cell relative to speed of diffusive transport; $\textrm{Pe}\approx 1--10$\\
 $K = \frac{k R (2 h)^2}{v_0}$ \dotfill &\textrm{Relative speed of reaction compared to motility; $K = O(100)$}\\
$\chi = \frac{\gamma}{2 h \alpha R}$ \dotfill &\textrm{Relative strength of tension vs actomyosin; $\chi \approx 0.2$}\\
 $\rhoz = \frac{\beta}{2 h \alpha}$  \dotfill &\textrm{Rescaled contractility; $\rhoz$ is the value of $\rho$ such that actomyosin force is zero, $0 \le \rhoz \le 1$}\\
 $C = \frac{N_{\textrm{tot}}}{2h R^2}$ \dotfill&\textrm{Rescaled total amount of $\rho$}\\
 $m$ \; \dotfill  &\textrm{Reaction parameter}\\
 $\epsilon = \tilde{\epsilon}/R$ \dotfill &\textrm{Rescaled interface size} \\
\hline
\end{tabular}
\caption{Table of unitless parameters}
\label{tab:unitless}
\end{table}

\section{Parameter estimation}\label{Section_ParaEsti}

Many of our parameters can be estimated well, or at least constrained, by using experimental data; other parameters may only be varied over a narrow range in order for our cell to effectively crawl.  

We are interested in modeling the crawling of keratocytes and other fast-moving cells \cite{verkhovsky1999self,keren2008mechanism,barnhart2011adhesion}, where cell speeds are in the range of $0.1 - 0.2 $ \um/s.  We will thus take $v_0 \approx 0.1$\um/s.  Keratocytes typically cover areas of around $30 \um \times 15 \um$ \cite{keren2008biophysical}, so we will assume an initial size scale of $R \approx 10 \um$.  In the model of cell polarity we use, we describe a Rho GTPase diffusing in the cell membrane with diffusion coefficient $D_\rho$; typical membrane protein diffusion coefficients of these Rho GTPases are of the order of $0.1 \um^2$/s \cite{postma2004chemotaxis}, though of course there may be some variation in this.  With these estimates, we expect $\textrm{Pe} = v_0 R /D_\rho$ to take on values ranging from $1$ to $10$, depending on the precise speed and diffusion coefficients involved.  We will often report parameters in terms of the inverse Peclet number, $\textrm{Pe}^{-1}$, which enters \eq \ref{eq:rho} as an effective diffusion coefficient in our units.  

The kinetic timescale of the Rho GTPases is expected to be on the order of seconds \cite{mori2008wave,sako2000single}; we will therefore set the rate $k (2h)^2$ to be of the order of $1/s$; this allows us to estimate $K = O(100)$.  

It is slightly more difficult to estimate $\chi = \frac{\gamma}{2 h \alpha R} = \frac{\gamma}{v_0 \tau R}$, as we need to determine the effective friction coefficient $\tau$ relating the force per unit length on the cell boundary to its velocity.  This friction is not simple, as it arises from a combination of hydrodynamic effects between the cell membrane and the substrate and friction from breaking adhesions with the surface; we do not know a convincing first-principles estimate of this value.  In \cite{shao2010computational}, a value of $\tau = 2.62 \, \textrm{pN} \textrm{s}/ \um^2$ was found to create a reasonable cell shape.  The tension on the membrane is estimated to be of the order $\gamma \approx 1 pN$ \cite{shao2010computational}, setting $\chi \approx 0.2$.  

$\rhoz$ is a rescaled contractility of the cell, measuring the ratio of forces driving contraction ($\beta$) to those driving protrusion ($2h \alpha$), and does not have a natural scale.  However, for the front of the cell, where $\rho \approx 1$, to protrude, we must have $\rhoz < 1$; we must also have $\rhoz > 0$ for the back of the cell, where $\rho \approx 0$, to contract, see \eq \ref{eq:pfu}.  

The values of $C$ and $m$ are constrained by the requirement that the cell be able to polarize.  These requirements include that \cite{mori2008wave,jilkine2009wave}:
\begin{align}
\frac{|\Omega|}{m} \le C \le \frac{m+1}{m}|\Omega| \label{eq:constraint}
\end{align}
where $|\Omega|$ is the area of the cell. Clearly as $m\to 0$, the cell will not be able to polarize unless $C$ is very carefully tuned.  We choose $m = 1/2$ and $C = 6$ throughout this paper, which we have found allows cells to polarize within a reasonable range of cell sizes.  We note that the asymptotic results in \cite{mori2008wave,jilkine2009wave} from which \eq \ref{eq:constraint} is derived are only completely valid for a stationary cell.  However, we have found similar transitions between polarized and unpolarized states in moving cells \cite{camley2013periodic}.

\section{Cell behavior}

\begin{figure}[t]
\includegraphics[width=80mm]{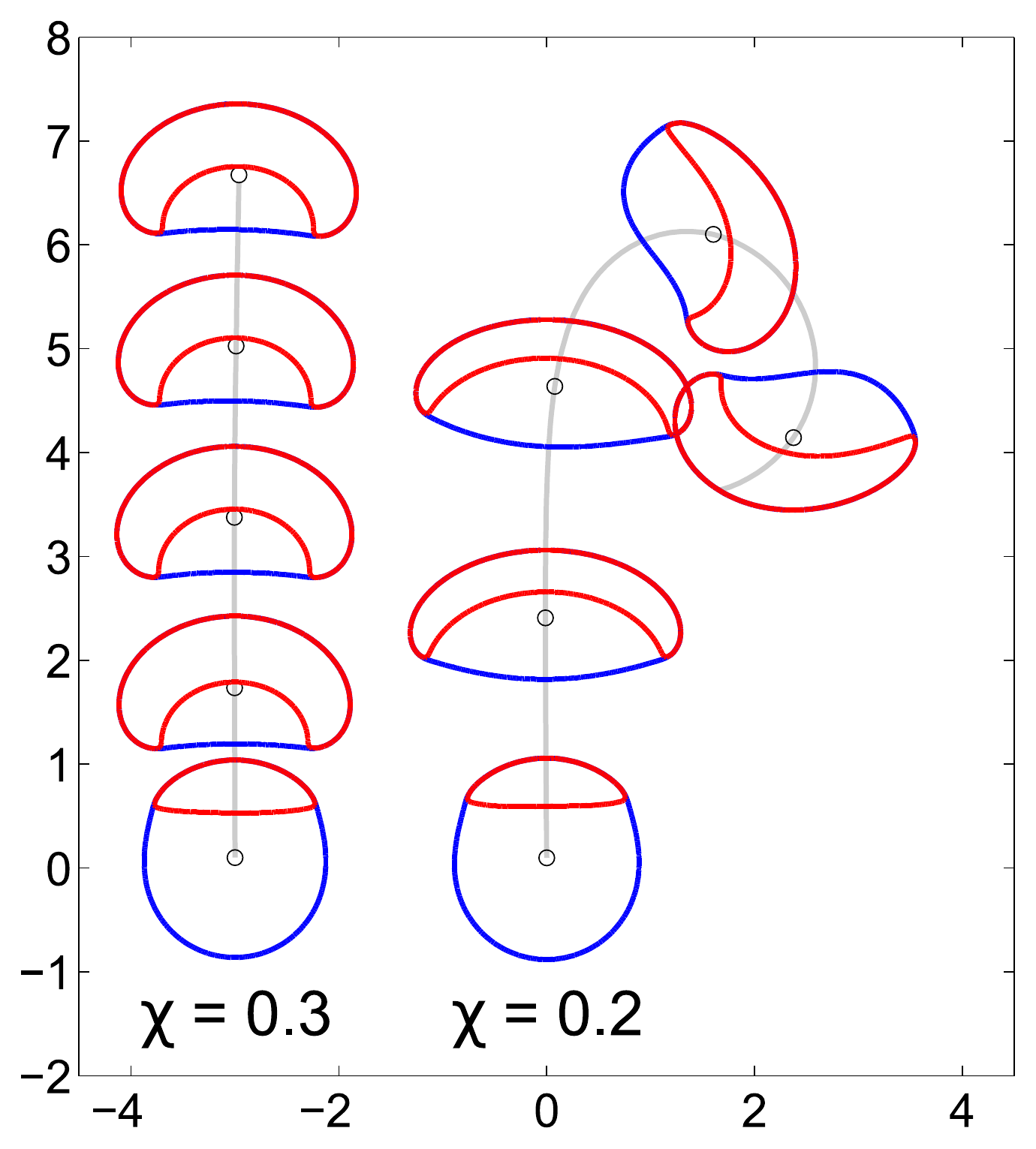}
\caption{Two characteristic trajectories of a crawling cell.  
Left: a crawling cell with a higher unitless tension ($\chi = 0.3$) maintains a persistent trajectory. Right: a crawling cell undergoes a turning instability and transitions to a circular trajectory when unitless tension is smaller ($\chi = 0.2$).  The snapshots are taken at $t = 0.2, 5.2, 10.2, \cdots, 25.2$. In each snapshot, the blue line indicates the cell boundary, i.e. $\phi = 1/2$, and the red line indicates the cell front - the half-maximum contour of the Rho GTPase $\rho$.  
$\rhoz = 0.4, \text{Pe}^{-1} = 0.30, K = 500, C = 6, m = 0.5, \epsilon = 0.1$ in these simulations. }  \label{straight}
\end{figure}

We numerically evaluate Eqs. \ref{eq:pfu}-\ref{eq:rhocyt} by a semi-implicit Fourier spectral method; see Appendix \ref{AppendixA} for numerical details.  We find that this simple model supports both straight and circular trajectories (\fig \ref{straight}).  Initially the cell shape is taken to be circular, and we choose the $\rho$-distribution to be polarized, $\rho = 0.8$ in the front half and zero in the rear, though with a random noise added on top (Appendix \ref{AppendixA}).  Due to this $\rho$-polarization, the front half of the cell is pushed out while the rear half is contracted, deforming the cell.  After nearly $t = 5$, the crawling cell with $\chi = 0.3$ reaches an equilibrium shape and undergoes a straight trajectory, while the one with $\chi = 0.2$ undergoes a turning instability and transitions to a circular trajectory at $t\approx 15$.  The straight trajectories resemble the highly persistent, half-moon shape of crawling keratocytes \cite{verkhovsky1999self,keren2008mechanism,barnhart2011adhesion}.  The turning behavior is equally likely to occur in either direction, and depends on the noise in the initial conditions; larger noise can accelerate turning, while states with zero initial noise can proceed for a very long time without turning.  

While we show only cells that effectively crawl in \fig \ref{straight}, we also note that even initially polarized cells may become depolarized, with $\rho$ becoming uniform over the cell.  This occurs at larger tensions than we plot here, in which the cell cannot effectively push the membrane out, and becomes too small to develop polarization.
This corresponds to violations of the constraints in \eq \ref{eq:constraint} in which wave-pinning fails \cite{mori2008wave,jilkine2009wave}, and only homogeneous solutions to the reaction-diffusion equations are possible.

\section{Transition to circular trajectories}

\begin{figure}[ht]
\includegraphics[width=90mm]{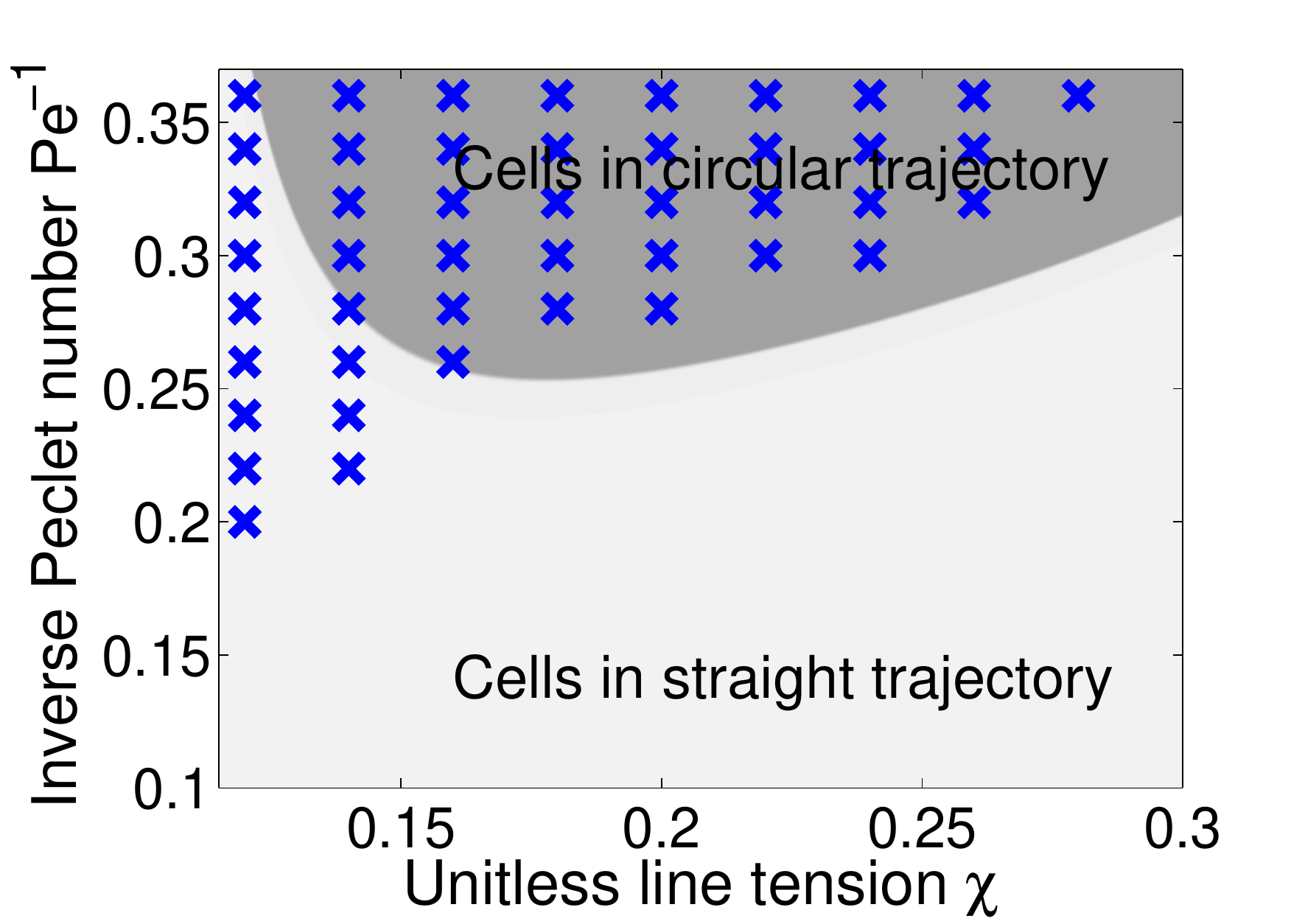}
\caption{(Color online) Cells turn at low $\chi$ and high $Pe^{-1}$.  Blue crosses show simulations where cells develop a circular trajectory.  Points are sampled from $\chi\in[0.12,0.3]$ and $\text{Pe}^{-1}\in [0.1,0.36]$ with sample grid 0.02. Other parameters are fixed as $\rhoz = 0.4, K = 500, C = 6, m = 0.5, \epsilon = 0.1$.  The background phase diagram is a sharp-interface prediction, where the darker gray corresponds to turning cells.  The theory is detailed in Section \ref{subsection:predicted_phase_diagram}.  Simulations are run until $t = 40$; if a cell has not turned by this point, we treat it as a stable one with straight trajectory. }  \label{chi_PeInv}
\end{figure}

What parameters control the transition from straight trajectories to circular ones? We show a $\chi$-$\text{Pe}^{-1}$ phase diagram  in \fig \ref{chi_PeInv}; in this phase diagram, blue crosses indicate the parameters at which we have observed cells turning and following circular trajectories.  In general, the straight trajectory is stabilized by increasing the unitless tension $\chi$ on the cell and decreasing the unitless diffusion coefficient $\text{Pe}^{-1}$ of the diffusing molecule $\rho$. 

\begin{figure}[t]
\includegraphics[height=60mm]{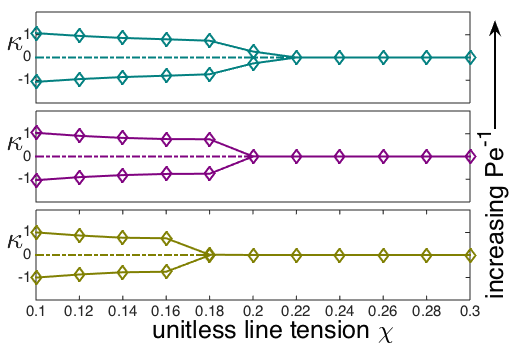}
\caption{(Color online) Curvature $\kappa$ of the circular-trajectory solutions of Eqs. \ref{eq:pfu}-\ref{eq:rhocyt} as a function of tension $\chi$. We plot these bifurcation diagrams for three values of the inverse Peclet number, $\text{Pe}^{-1} = 0.26, 0.28, 0.30$ from bottom to top.  Dashed lines indicate that the straight crawling cell trajectory is unstable under small perturbation.}  \label{radius_bifurcation_chi}
\end{figure}

Additionally, for a crawling cell in circular trajectory, the curvature $\kappa$ of the trajectory is affected by $\chi$ and $\text{Pe}^{-1}$.  We show a  bifurcation diagram of $\kappa$ as a function of $\chi$ for three values of $\textrm{Pe}^{-1}$ in \fig \ref{radius_bifurcation_chi}. Curvature $\kappa = 0$ indicates the straight trajectory of the crawling cell.  We see the existence of a supercritical pitchfork bifurcation point for $\chi$ below which the crawling cell tends to undergo circular motion (the up-down symmetry in the bifurcation diagram is due to the fact that the cell can turn to clockwise circular motion or counterclockwise circular motion  with equal probability), and after which the cell becomes stable in a straight trajectory. The dashed line indicates that for small value of $\chi$, the straightly crawling cell is unstable in the sense that it will turn to a circular motion under small perturbation in the system. As $\text{Pe}^{-1}$ increases, the bifurcation value of $\chi$ becomes smaller, which is also observed in the phase digram \fig \ref{chi_PeInv}.

\begin{figure}[ht]
\includegraphics[width=90mm]{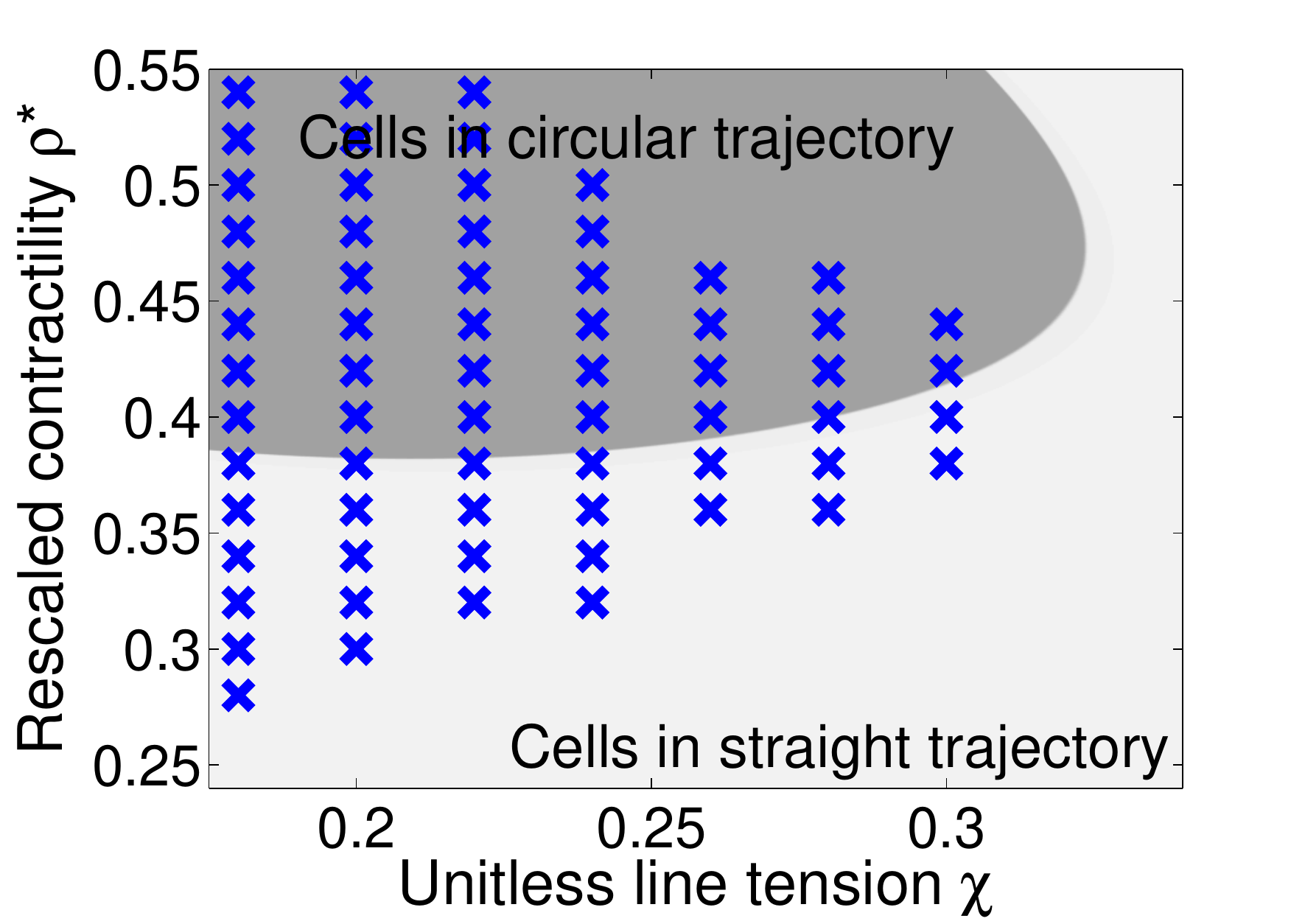}
\caption{(Color online) Phase diagram for cell turning as a function of tension $\chi$ and rescaled contractility $\rhoz$.  Blue crosses show simulations where cells develop a circular trajectory.  Points are sampled from $\chi\in[0.18,0.4]$ and $\rhoz\in [0.28,0.58]$ with sample grid size of 0.02. Other parameters are fixed as $\text{Pe}^{-1} = 0.3, K = 500, C = 6, m = 0.5, \epsilon = 0.1$.  Increasing the line tension $\chi$ tends to stabilize the cell; while increasing $\rhoz$ first destabilizes the straight trajectory and then restabilizes it.  The background phase diagram is a sharp-interface prediction, where the darker gray corresponds to turning cells.  The theory is detailed in Section \ref{subsection:predicted_phase_diagram}. Simulations are run until $t = 40$; if a cell has not turned by this point, we treat it as a stable one with straight trajectory.}  \label{fig:chi_rho0}
\end{figure}

We plot a phase diagram for cell turning as a function of tension $\chi$ and rescaled contractility $\rhoz$ in \fig \ref{fig:chi_rho0}; again, blue crosses indicate where the cell turns to a circular trajectory.  Surprisingly, as we vary $\rhoz$ we observe \textit{reentry}, as increasing $\rhoz$ first destabilizes the straight trajectory and then subsequently restabilizes it.  

The phase diagrams in both \fig \ref{fig:chi_rho0} and \fig \ref{chi_PeInv} show both simulation results (blue crosses) and a theoretical approximation of the phase diagram (dark/light gray coloring).  We will now introduce this theoretical analysis, which will provide some intuition about the influence of $\chi$, $\textrm{Pe}^{-1}$, and $\rhoz$.  

\section{Origin of turning instability: analytical estimate of phase diagram}

\begin{figure}[ht]
\includegraphics[width=90mm]{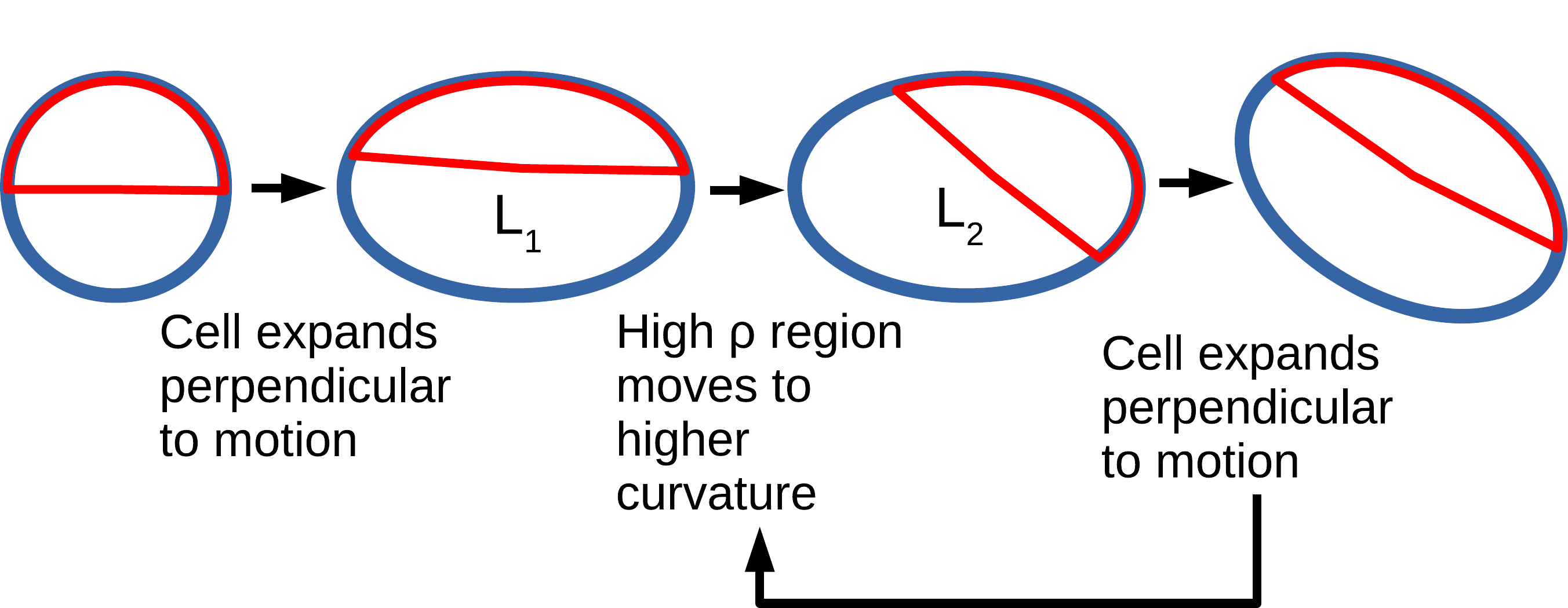}
\caption{(Color online) Proposed schematic diagram for turning instability. Cell polarization leads to cell deformation (seen in \fig \ref{straight}); this leads to polarity reorientation to minimize the length of the interior $\rho$ interface, as $L_2 < L_1$; this new $\rho$ polarization will alter the cell shape. This is a potential biomechanical mechanism for the generic instability proposed in \cite{ohta2009deformable}.}  \label{diagram}
\end{figure}

Why does turning occur? We argue that the turning instability and circular trajectory arise from the coupling between the reaction-diffusion mechanism and the cell shape in \eq \ref{eq:rho}, in a scheme illustrated in \fig \ref{diagram}. The reaction-diffusion mechanism tends to minimize the length of the interface between regions of high $\rho$ and low $\rho$. For this reason,  the high $\rho$ regions are attracted to high curvature \cite{vanderlei2011multiscale, jilkine2009wave}. As the cell is deformed, widening in the direction perpendicular to cell motion, the cell front becomes a local curvature minimum (\fig \ref{straight}), and the reaction-diffusion mechanism will, if the cell shape is fixed, re-orient the cell polarization to the high curvature region. The rate of this destabilizing process is controlled by the effective diffusion coefficient $\text{Pe}^{-1}$. 

Given this instability, why can the cell maintain a straight trajectory if $\text{Pe}^{-1}$ is low enough or $\chi$ high enough? Even if the polarity is linearly unstable, the straight trajectory may be rescued by the cell's ability to adapt its shape. If the cell shape immediately reorients to any change in polarity, the cell cannot turn, and the straight trajectory is stabilized. This explain the importance of $\chi$, as the dynamics of the cell shape strongly depend on $\chi$.  We would then naturally expect increasing $\chi$ to stabilize the cell, and increasing $\text{Pe}^{-1}$ to destabilize it, as seen in \fig \ref{chi_PeInv}.  However, this intuition does not immediately explain the effect of $\rhoz$ in \fig \ref{fig:chi_rho0}, which we will find to occur because of the influence of $\rhoz$ on cell shape.

We have been able to qualitatively, but not quantitatively reproduce the phase diagrams sketched in \fig \ref{chi_PeInv} and \fig \ref{fig:chi_rho0}, including the reentry, with a calculation based on this argument.  In Sec. \ref{section:rho}, we show that the $\rho$ dynamics respond to the cell shape, and show that this instability is slower as Pe increases, but also depends on the cell shape.  In Sec. \ref{section:interface}, we study the dynamics of how the cell shape is controlled by the distribution of $\rho$, and show that the speed at which a cell relaxes to its new shape is proportional to $\chi$.  In Sec. \ref{subsection:predicted_phase_diagram}, we combine these results to predict the phase diagram of cells as a function of the effective cell tension $\chi$, the Peclet number, and the rescaled contractility $\rhoz$.  

\subsection{Dynamics of Rho GTPase within a fixed geometry}\label{section:rho}

\begin{figure}[ht]
\includegraphics[width=90mm]{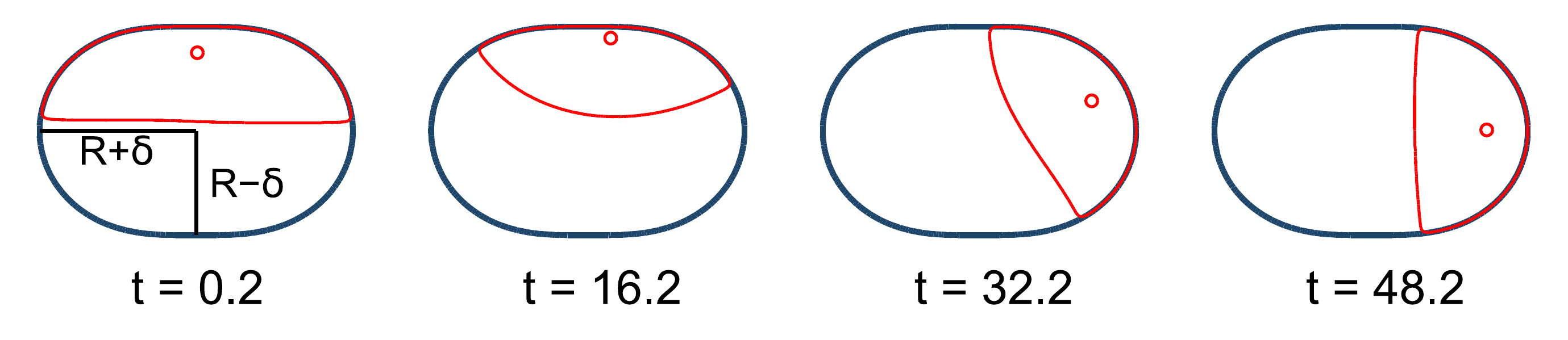}
\caption{(Color online) Simulation of $\rho$ dynamics in a fixed geometry shows that the high-$\rho$ cell front region is attracted to regions of higher curvature, as noted in \cite{jilkine2009wave,vanderlei2011multiscale}.  As a result, in a static near-elliptical geometry, the polarity will tend to move toward the narrow end.  (Cells are rotated to show their elongated dimension along $x$.)  }  \label{rotate}
\end{figure}

Simulating the reaction-diffusion dynamics of \eq \ref{eq:rho} in a fixed, non-moving cell shows that the reaction-diffusion mechanism is sensitive to cell shape (\fig \ref{rotate}).  In particular, we find that the cell ``front'' (region of high $\rho$) rotates to point toward the higher-curvature region of the cell.  This is consistent with the idea that the reaction-diffusion dynamics of \eq \ref{eq:rho} serve to minimize the interface between high $\rho$ and low $\rho$.  This behavior has been noted before \cite{jilkine2009wave,vanderlei2011multiscale}; see also a brief discussion of this point in \cite{wolgemuth2011redundant}.  

We characterize the kinetics of the instability of large $\rho$ moving to higher curvature, and how it depends on the cell shape and Peclet number.  We will look at the dynamics of $\rho$ in a cell with a fixed shape,
\begin{align*}
R(\theta) = R_0 - \delta \cos 2\theta
\end{align*}
where $\theta$ is, as usual, the angle counter-clockwise from the $x$ axis.  To compute the evolution in $\rho$ in a fixed cell shape, we solve only \eq \ref{eq:rho} with a fixed $\phi(\rb) = \frac{1}{2}\left[1+\tanh\left(3 r_s /\epsilon\right)\right]$, where $r_s$ is the (signed) distance from the curve $R(\theta)$.

\begin{figure}[ht!]
\includegraphics[height=70mm]{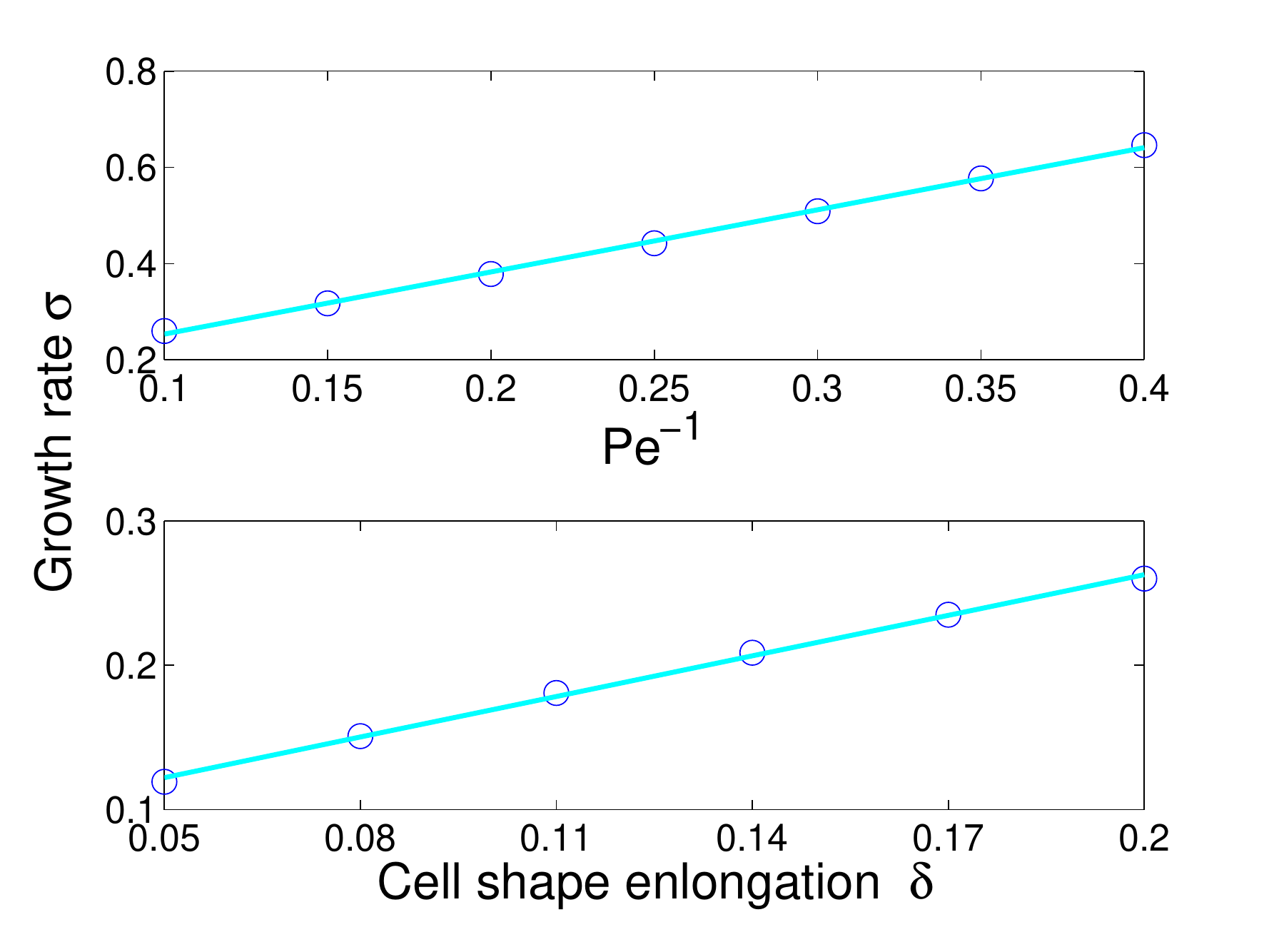}
\caption{The growth rate $\sigma$ of the reorientational $\rho$ instability in a fixed cell geometry depends linearly on both the cell deformation $\delta$ and $\text{Pe}^{-1}$.   The parameter $\delta = 0.3$ is fixed in the top subfigure, while in the bottom subfigure, $\text{Pe}^{-1} = 0.1$ is fixed. Other parameters take their default values. }  \label{fig:sigma_plot}
\end{figure}

The instability in \fig \ref{rotate} is a linear instability.  We find that if the distribution of $\rho(\theta)$ is initially centered near $\theta = 0$, we find an exponential increase of the center of mass of $\rho(\theta)$ with time, $\theta_{\rho} = \theta_{\rho}(0) e^{\sigma t}$.  (Here, we define $\theta_\rho$ to be the angle to the center of mass of the distribution $\rho(\rb)$, $\vb{r}_{\rho} = \int d^2 r \vb{r} \rho(\rb) \phi(\rb) / \int d^2r \rho(\rb) \phi(\rb)$.)
If this instability is driven by the mean curvature flow identified by Ref. \onlinecite{jilkine2009wave}, we would expect that $\sigma \sim \text{Pe}^{-1}$.  We have confirmed this numerically for the parameters we have studied (\fig \ref{fig:sigma_plot}).

We also expect that as $\delta\rightarrow 0$, the instability should vanish as a perfect circle has no shape asymmetry.  We find, consistently with this intuition, that $\sigma \sim \delta$ at small $|\delta|$ (\fig \ref{fig:sigma_plot}).  Based on these results, we hypothesize that $\sigma = b\delta \text{Pe}^{-1}$ with $b$ a constant, i.e.
\begin{align}
\dfrac{d}{dt} \theta_{\rho} = b \delta \text{Pe}^{-1} \theta_\rho \label{eq:sigma_inst}
\end{align}
Though this is reasonable at small $\delta$, we would not expect it to necessarily generalize to larger aspect ratios.  In addition, in a more complex shape, $\sigma$ may depend on higher Fourier modes in the cell shape.

\subsection{Dynamics of cell shape response to $\rho$} \label{section:interface}

\subsubsection{Sharp interface theory} 

In the absence of any driving forces, or $\rho = \rhoz$, the phase field \eq \ref{eq:pfu} is simply an Allen-Cahn equation. In the sharp interface limit of $\epsilon\rightarrow0$, the interface evolves with a normal velocity $v_n = -\chi\kappa$ with $\kappa$ the interface curvature \cite{elder2001sharp} (we note this is distinct from the curvature of the trajectory, which we also have labeled $\kappa$ above, but is not addressed in this section). The added forcing terms correspond to the interface being advected with velocity $\vb{v} = (\rho-\rhoz)\hat{\vb{n}}$ where $\hat{\vb{n}} = -\nabla\phi/|\nabla{\phi}|$ is the outward-pointing normal. Since this term varies smoothly across the boundary, we can get the normal interface velocity by adding $\rho-\rhoz$ directly to the curvature-driven relaxation velocity $-\chi\kappa$ \cite{han2011comprehensive}. Then the normal velocity of the cell boundary in arc-length $s$ should be:
\begin{align}\label{eq:sharp_interface_limit_normal_velocity}
v_n(s) = (\rho-\rhoz) - \chi\kappa(s)
\end{align}
When the cell takes on a steady shape, the normal velocity must satisfy:
\begin{align}\label{average_vn_zero}
\int ds v_n(s) = 0.
\end{align}
where the integral is over the entire cell boundary.  

\subsubsection{Steady cell shape in quasi-circular Fourier modes} \label{sec:fourier}

Let us describe a cell with boundary given by the function $R(\theta,t)$, where the angle $\theta$ is with respect to the $x$ axis . We can expand $R$ in Fourier modes:
\begin{align}\label{Fourier_expansion}
R(\theta,t) = R_0 + \sumt c_n(t) e^{in\theta}.
\end{align}
Here we assume that the cell shape is close to circular, $c_n/R_0\ll 1$, and we use $\sumt$ to denote $\sum_{\neq0,\pm1}$. The $n = \pm1$ mode is excluded because it corresponds to the cell translational motion, which we include by assuming that the cell is initially traveling with a constant speed of $\vb{v}$.  We have also excluded the $n = 0$ size expansion mode; we find, in a numerically exact solution of \eq \ref{eq:sharp_interface_limit_normal_velocity}, that there are many distinct solutions corresponding to different cell perimeters.  Selecting $R_0$ or equivalently $c_0$ chooses which of these steady state solutions we observe, and will be done later by setting the cell perimeter $L \approx 2\pi R_0$.  

The normal velocity at an angle $\theta$ is \cite{ohta2009deformation}
\begin{align}
v_n(\theta) = \vb{v}\cdot\hat{\vb{n}} + \dfrac{R}{\sqrt{R^2 + R'^2}}\dfrac{dR}{dt}
\end{align}
Up to the first order of the deviations, we have
\begin{align}
\kappa(\theta,t ) &= \dfrac{1}{R_0} + \dfrac{1}{R_0^2}\sumt(n^2-1)c_n(t) e^{in\theta}\\
v_n(\theta,t) & = \vb{v} \cdot \vb{\hat{n}} +  \sumt \dfrac{d}{dt} c_n(t) e^{in\theta}
\end{align}
Expanding \eq \ref{eq:sharp_interface_limit_normal_velocity} into Fourier modes, we find (assuming the cell has velocity $\vb{v} = v\vb{\hat{x}}$),
\begin{align}\label{eq:cn}
\dfrac{d}{dt} c_n(t) &= \rho_n - \dfrac{\chi}{R_0^2}(n^2-1) c_n  \\
\nonumber &+ \frac{v}{2 R_0} \bigg[(n+1) c_{n+1} - (n-1) c_{n-1}\bigg],\quad n \neq 0,\pm 1
\end{align}
We will treat $c_{0}$ and $c_{\pm 1}$ as zero when they arise from the term proportional to $v$ here, and
\begin{align}
\rho_n = \dfrac{1}{2\pi} \int_{-\pi}^{\pi} d\theta \left[\rho(\theta)-\rhoz\right] e^{-in\theta}
\end{align}
is the Fourier transform of the protrusion strength -- how far the Rho GTPase protein $\rho$ exceeds the critical value $\rhoz$. 

The equations of motion for the Fourier modes $c_n(t)$ (\eq \ref{eq:cn}) depend on the velocity of the cell, $\vb{v}$.  This velocity can be found, following \cite{ohta2009deformation}, as 
\begin{align}
\vb{v} &= \frac{1}{A}\int ds \vb{R}(s) v_n(s)\\
\nonumber	   &\approx \frac{\hat{\vb{x}}}{A}\int_{0}^{2\pi} d\theta \left(R_0^2 + 2 R_0 \sum_{n} c_n e^{in\theta} \right) \times \\
\nonumber & \hspace{30mm} \left[\rho(\theta)-\rhoz - \chi K(\theta) \right] \\
	   &= \hat{\vb{x}}\left[\rho_1 + \rho_{-1} + \frac{2}{R_0} \sum_n c_n \left( \rho_{-n+1} + \rho_{-n-1} \right) \right] \label{eq:velocity_rho}
\end{align}
where $\vb{R}(s)$ is the vector pointing from the cell's center of mass to the element at arclength $s$, $A \approx \pi R_0^2$ is the cell area, and the approximation is true for small deformations.

We have seen from our simulations and the analysis of \cite{mori2008wave, mori2011asymptotic} that $\rho$ has a sharp interface between values $\rho^+ = m\rho_{\text{cyt}}=1$ and $\rho^-=0$, which are controlled by the details of the reaction term \eq \ref{eq:rho}. We thus assume a simple form for $\rho(\theta)$:
\begin{align}
\rho(\theta) = 
\begin{cases}
\rho^+,\quad &|\theta|\le \theta^+/2\\
0, \quad &\text{otherwise}
\end{cases}
\end{align}
with $\theta^+$ indicating the angle of the cell over which $\rho$ is equal to $\rho^+$, with $0<\theta^+<2\pi$.  
With this form,
\begin{align}\label{eq:rhon}
\rho_n = 
  \dfrac{\rho^+}{n\pi}\sin\left(\dfrac{n\theta^+}{2}\right).
\end{align}
Importantly, we can find $\theta^+$ without explicitly solving the reaction-diffusion equations.  Integrating \eq \ref{eq:sharp_interface_limit_normal_velocity} over the arc-length and using \eq \ref{average_vn_zero}, we find a relationship between $\rho$, the cell shape, and $\chi$ that is required for there to be a steady state shape,
\begin{align}\label{eq:norm_velocity_integrate_over}
2\pi\chi = \int_0^L ds (\rho-\rhoz),
\end{align}
where $L\approx 2\pi R_0 + O(c_n)$ is the cell perimeter. The integral over the arc length $s$ can be cast into one over the angle $\theta$:
\begin{align}
ds &= \sqrt{R(\theta)^2 + R'(\theta)^2} d\theta \label{eq:contour} \\
&= \left(R_0 + \sumt c_n e^{in\theta} + O(c_n^2)\right) d\theta,
\end{align}
then the \eq \ref{eq:norm_velocity_integrate_over} becomes, up to linear order in $c_n$, 
\begin{align}
\nonumber 2\pi\chi &= \int_{-\pi}^{\pi} d\theta(\rho-\rho^*)R_0  + \sumt c_n \rho_{-n}\\
& = R_0\rho^+\theta^+ - 2\pi R_0\rho^*   + 2\rho^+ \sumt\dfrac{c_n}{n}\sin\left(\dfrac{n\theta^+}{2}\right). \label{eq:theta+}
\end{align}
This equation is a link between cell shape and $\theta^+$ at steady state.  Expanding $\theta^+$ in $c_n$ as $\theta^+ = \theta^+_0 + \theta^+_1 + \cdots$, where we assume $\theta^+_1$ is $O(c_n)$, we find 
\begin{align}\label{eq:theta+0}
\theta^+_0 = \dfrac{2\pi(\chi + R_0\rhoz)}{R_0\rho^+}.
\end{align}
and
\begin{align}
\theta^+_1 = - \dfrac{2}{R_0} \sumt \dfrac{c_n}{n}\sin\left(\dfrac{n\theta^+_0}{2}\right).
\end{align}

Then the Fourier modes of $\rho$ in \eq \ref{eq:rhon} becomes, to linear order again,
\begin{align}
\nonumber \rho_n = \dfrac{\rho^+}{n\pi}\sin\left(\dfrac{n\theta^+_0}{2}\right) +  \dfrac{\rho^+}{2\pi}\cos\left(\dfrac{n\theta^+_0}{2}\right) \theta_1^+.
\end{align}

If we look for the steady state of $c_n$, which we will write $\css{n}$, we find that, using \eq \ref{eq:cn},
\begin{align}
\dfrac{\chi}{R_0^2}(n^2-1)\css{n} - \frac{v}{2 R_0} \left[ (n+1) \css{n+1} - (n-1)\css{n-1}\right] \\
= \rho_n(\css{n}) \nonumber
\end{align}
which can be re-written as a simple matrix multiplication,
\begin{align}\label{eq:cnss}
\sum_{m\neq 0 \pm1} A_{nm} \css{m} = g_n
\end{align}
where
\begin{align}
\nonumber A_{nm} &= \dfrac{\chi}{R_0^2}(n^2-1) \delta_{nm} - \frac{m v}{2R_0}\left[\delta_{n+1,m} - \delta_{n-1,m}\right] + B_{nm}\\
\nonumber B_{n,m} &= \frac{\rho^+}{m\pi R_0} \cos\left(\dfrac{n \theta_0^+}{2}\right) \sin\left(\dfrac{m \theta_0^+}{2}\right) \\
\nonumber g_n &= \frac{\rho^+}{n \pi} \sin\left(\dfrac{n \theta_0^+}{2}\right)
\end{align}
where $\delta_{mn}$ is the Kronecker delta. Because $v$ multiplies terms of order $c_n$, we can approximate it by 
\begin{align}
v &\approx \rho_1 + \rho_{-1} \\
&\approx (2 \rho^+/\pi) \sin\left(\theta_0^+/2\right)
\end{align} to zeroth order in $c_n$ (\eq \ref{eq:velocity_rho}).

\eq \ref{eq:cnss} may be solved to reconstruct $c_n^{\text{s.s.}}$, and therefore $R(\theta)$, by truncating to finite number $N_{\textrm{max}}$ of Fourier modes, $n \in (-N_{\textrm{max}},\cdots,-3,-2,2,3,\cdots,N_{\text{max}})$. 
If we only take the $n=\pm 2$ modes, the answer is relatively simple,
\begin{align}\label{c2ss}
\css{\pm 2} &= \dfrac{\rho^+}{2\pi}\sin\theta^+_0\left( \dfrac{3\chi}{R_0^2} + \dfrac{\rho^+\sin 2\theta^+_0}{2\pi R_0}\right)^{-1} 
\end{align}

This model, with the assumptions we have made, is straightforward to solve.  However, it is not numerically exact because of the assumption of quasi-circularity, $c_n/R_0\ll 1$.  We compare the Fourier series shapes with a sharp-interface determination of the cell shape that does not require assuming $c_n/R_0 \ll 1$ in Appendix \ref{subsection:exact_calculation}.

\subsubsection{Dynamics of Perturbation from Steady-state Shape}

In calculating the steady-state shape above, we have assumed that the cell travels at a steady velocity in the $\hat{\textbf{x}}$ direction. Our simulations show that turning begins from a near-steady-state shape.  To study this linear instability, we will calculate how the cell shape relaxes if we slightly change the distribution of $\rho(\theta)$.  If the orientation of $\rho(\theta)$ changes by a small angle $q$, this is exactly the same as if we slightly rotate our cell shape away from the $x$-axis, and see how it relaxes.  We assume that this process is dominated by the dynamics of the lowest mode, $n = \pm 2$; we will thus look at the dynamics of $c_2$ when it takes on the form $c_2 =  c_2^{\text{s.s.}} e^{iq(t)}$ with $q(t)$ small, and neglect all other modes $|n|>2$.  We will also assume that $\theta^+$ and $v$ do not depend on $q$.  With these assumptions, we find from \eq \ref{eq:cn} that, to linear order in $q$,
\begin{align}
\dfrac{d}{dt} q = - \dfrac{3\chi}{R_0^2}q. \label{eq:qrelax}
\end{align}

We note that because we have limited ourselves to the $n = \pm 2$ modes, the relaxation dynamics of this rotation do not depend on the cell's velocity $v$. 

\subsection{Predicted Phase Diagram}\label{subsection:predicted_phase_diagram}

We can now predict when the cell should be stable or unstable.  Small perturbations of cell shape away from the direction of polarity relax with a rate $3\chi/R_0^2$, as shown in \eq \ref{eq:qrelax}.  We also found numerically that, in a fixed cell shape with a distortion size of $\delta$, the front of the cell will move toward the narrow end of the cell with a rate $\sigma = b\delta \text{Pe}^{-1}$ (\fig \ref{fig:sigma_plot} and \eq \ref{eq:sigma_inst}).  Combining these results will show when the linear cell motion remains stable.

In our earlier results, computing the shape relaxation, we assumed that the initial direction of polarity was $\theta_{\rho} = 0$, but this is not necessary. Similarly in computing the instability of $\rho$ in a static cell shape, we assumed a stationary shape that is narrowed along the $x$-axis, but we can rotate to consider the shape relative to an arbitrary axis to get equivalent results. We can then generalize our above results to
\begin{align}
\dfrac{d}{dt} q &= -\dfrac{3\chi}{R_0^2}(q-\theta_{\rho})\ ,\\
\dfrac{d}{dt}\theta_{\rho} &= \sigma(\theta_{\rho} - q)\ .
\end{align}
Combining these equations, we find that the linear stability of $\theta_{\rho} - q$, i.e. the difference between the direction of chemical polarity $\rho$, and the direction of shape polarity, $q$, is controlled by
\begin{align}
\dfrac{d}{dt} (\theta_{\rho} - q) = \left[ \sigma - \dfrac{3\chi}{R_0^2}   \right] (\theta_{\rho} - q),
\end{align}
from which we can see that when $\sigma < 3\chi/R_0^2$, we expect our straight-crawling cell to be stable to linear perturbations, and for $\sigma > 3\chi/R_0^2$ we expect it to turn. 

We established that $\sigma = b\delta \text{Pe}^{-1} = -2b c_2^{\text{s.s.}}\text{Pe}^{-1}$.  Using our simulations (\fig \ref{fig:sigma_plot}), we estimate $b \approx 9.3724$.  The only other crucial feature is the steady-state shape of the crawling cell $c_2^{\text{s.s.}}$, which we know by \eq \ref{c2ss} -- assuming once again that the cell shape is dominated by the lowest $n = \pm 2$ mode.  We then have the bifurcation relation for marginal stability:
\begin{align}
-\dfrac{b\rho^+}{\pi}\sin\theta^+_0\left( \dfrac{3\chi}{R_0^2} + \dfrac{\rho^+\sin 2\theta^+_0}{2\pi R_0}\right)^{-1} \text{Pe}^{-1} = \dfrac{3\chi}{R_0^2} \label{eq:bifurcation}
\end{align}
where $\rho^+ = 1, R_0 = 1$ and $\theta_0^+$ is set by \eq \ref{eq:theta+0}.

We show slices of this phase diagram in \fig \ref{chi_PeInv} and \fig \ref{fig:chi_rho0}. The phase diagrams we compute are only roughly accurate, as would be expected with the number of approximations that we have made.  However, we predict correctly both the order of magnitude of the transitions, and that there should be a {\it reentry} as $\rhoz$ decreases, where for both small $\rhoz$ and large $\rhoz$ there is stability (\fig \ref{fig:chi_rho0}). However the theory also predicts a reentry for small $\chi$; this has not been observed in the simulations.  This may be because as the tension $\chi$ becomes smaller, the cell shape becomes less and less quasi-circular, and our assumptions fail.

\section{Discussion}

We argue that the existence of turning and circular motion may be quite generic in cell motility of the sort we have studied here, with biochemical polarity mechanisms that create a single front.  Many other reaction-diffusion dynamics or other potential biochemical models of the cell's polarity \cite{jilkine2011comparison} may display the attraction to high curvature which drives the instability we discuss here. For instance, polarity driven by phase separation of two non-miscible species \cite{zamparo2015dynamic} would also tend to minimize the interface between these species.  Related mechanisms, including phase separation, and the constrained Allen-Cahn equation, are known to display instabilities similar to that of \fig \ref{rotate} within fixed geometries \cite{alikakos2000motion,stafford2001dynamics,alikakos2000mullins,marenduzzo2013phase}.  Turing patterns may also be reoriented by curvature, though in some reaction-diffusion systems, coupling to curvature can be overwhelmed by initial conditions \cite{orlandini2013domain,vandin2016curvature}.  We also note that the coupling between shape and protein dynamics has been emphasized recently in a Rho GTPase model wave pinning model applied to dendritic spines \cite{ramirez2015dendritic}, and cell shape-biochemistry interactions have been observed in a broad range of models and experiments \cite{camley2013periodic,holmes2012modelling,meyers2006potential}.  In general, we would expect any mechanism for $\rho$ that displays an effective line tension at the region between high and low $\rho$ to be able to generate turning instabilities of the type studied in our paper.  

In addition, Ohta and Ohkuma \cite{ohta2009deformable} have argued from a simple model proposed on symmetry grounds that the transition to circular motion is a generic property of active deformable particles, as long as there is a coupling between particle shape and particle polarity.  Our model provides a possible example of this coupling in the context of cellular motility.  However, the details of our mechanical model shows that generic models of this sort (e.g. \cite{ohta2016simple,tarama2013oscillatory,tarama2012spinning,menzel2012soft,hiraiwa2010dynamics}) can conceal surprises like reentry -- it is not at all straightforward to map physical properties of cells into the effective parameters. In particular, because the destabilizing effect of the reaction-diffusion mechanism depends on the steady-state cell shape, any parameter that controls cell shape can alter the stability diagram, and cell shape may not be a simple monotonic function of changing physical parameters.  

Cell turning has been studied in other models \cite{maree2006polarization,mogilner2010actin}, though primarily in a response to an altered stimulus -- e.g. the rotation of a chemoattractant gradient or an actin asymmetry.  This is in contrast to our example, where turning occurs spontaneously.  However, we do note that Ref. \onlinecite{maree2006polarization}, observes a drifting behavior which could be a transition into a very large-radius circular turn.  

We have also observed turning and circular motion in the full model of \cite{shao2012coupling, camley2013periodic}, which includes fluid flow, separate dynamics for myosin, and individual adhesions with stochastic transitions: see \fig S1 in the Supplementary Material of \cite{camley2013periodic}. We have found that decreasing $D_{\rho}$ ($D_a$ in \cite{camley2013periodic}) also tends to stabilize the cell in the more complex model. However, mechanical parameters do not have as straightforward an effect as studied in the simple model \eq \ref{eq:pfu}-\ref{eq:rhocyt}; in particular, we were unable to stabilize turning cells by straightforwardly increasing tension.  

How do our results on turning and circular motion compare with experiments on cell motility?  Recent work has shown that multi-lobed keratocytes undergo circular motion \cite{raynaud2016minimal}, but with a very different shape than the cells we simulate.  In addition, Gorelik and Gautreau have recently suggested that arpin \cite{dang2013inhibitory} may induce cells to turn by slowing them \cite{gorelik2015arp2}; this is consistent with our result that slowing the cell (or decreasing the Peclet number) can cause turning.  However, we emphasize that in other cell types turning is associated with different morphology and may not be controlled by the simple mechanism studied here \cite{liu2015linking}.  

We predict, based on our analysis, that when our mechanism applies, cell slowing will correspond with increased turning, but by a different mechanism than the speed-persistence relationship identified by \cite{maiuri2015actin}. Turning could potentially be prevented by reducing the membrane diffusion coefficient of polarity proteins on the surface of the cell, e.g. by increasing their binding to the cortex.  We also argue that cell shape is a crucial mediator of turning: wider cells would, in this mechanism, tend to be less stable.  Any interventions that alter cell tension, contractility at the cell rear, or strength of protrusiveness, and thereby alter cell shape may disrupt or induce turning.

\section{Conclusions}

In this paper, we have presented a simplified variant of a phase field cell motility model, extending our earlier work \cite{shao2010computational,camley2014polarity}.   We demonstrated that our model can support both straight and circular trajectories, with the circular trajectories occurring through a turning instability.  We have argued that this instability occurs because of the instability of the protein dynamics model we have adapted \cite{mori2008wave} that tends to orient proteins within the cell toward the cell's narrower ends.  When combination of the protein dynamics and cell shape dynamics leads to the cell widening, this may lead to destabilization of the straight trajectory.  Both our model and our simple theory suggest that the phase diagram of turning can be highly complex, with changing parameters having non-monotonic effects on the stability: we observe that increasing contractility first destabilizes and then restabilizes the cell's straight trajectory.  

\appendix

\section{Numerical Method} \label{AppendixA}

For the numerical method of the phase field model  (Eqs. \ref{eq:pfu}-\ref{eq:rhocyt}), we adopt the semi-implicit Fourier spectral method. 

Let us consider a rectangular domain in $\mathbb{R}^2$:
\[
\Omega = \{-L_x<x<L_x, -L_y<y<L_y\}
\]
and a periodic boundary condition is imposed for the problem. Let us discretize the spatial domain $\Omega$ by a rectangular mesh which is uniform in each direction as follows:
\[
(x_i,y_j) = (-L_x + i h_x, -L_y + j h_y)
\]
for $0\le i \le N_x$ and $0\le j \le N_y$, $h_x = 2L_x/N_x$ and $h_y = 2L_y/N_y$. Let $\phi_{ij} = \phi_{ij}(t)\approx \phi(x_i,y_j, t)$, $\rho_{ij} = \rho_{ij}(t)\approx \rho(x_i,y_j, t)$ denote the approximate solutions. Then the set of unknowns are
\[
\Phi = (\phi_{ij})_{0:N_x-1, 0:N_y-1}, U = (\rho_{ij})_{0:N_x-1, 0:N_y-1}.
\]
The Laplacian operator in the spectral space corresponds to the following spectrum
\[
\lambda_{ij} = -\lambda_x^2(i) - \lambda_y^2(j)
\]
with 
\begin{align*}
\lambda_x(i) = 
\begin{cases}
\frac{\pi}{L_x}i, &0\le i\le N_x/2,\\
\frac{\pi}{L_x}(N_x-i), &N_x/2<i\le N_x-1,
\end{cases}\\
\lambda_y(j) = 
\begin{cases}
\frac{\pi}{L_y}j, &0\le j\le N_y/2,\\
\frac{\pi}{L_y}(N_y-j), &N_y/2<j\le N_y-1.
\end{cases}
\end{align*}

For the \eq \ref{eq:pfu}, we can write it into the semi-implicit form:
\begin{align*}
\phi(\vb{x}, t&+\Delta t)-\chi\Delta t \nabla^2\phi(\vb{x},t+\Delta t) =\\
& \phi(\vb{x},t) + \Delta t(\rho-\rhoz)|\nabla \phi(\vb{x},t)| - \dfrac{\chi\Delta t}{\epsilon^2}G'(\phi(\vb{x},t)).
\end{align*}
By taking the fast Fourier transform of both sides of the above equation, we get
\[
(1 + \chi\Delta t \Lambda)\odot\hat{\Phi}(t+\Delta t) = \textbf{FFT}(R.H.S.)
\]
where $\Lambda = (\lambda_{ij})_{0:N_x-1, 0:N_y-1}$ and $\odot$ stands for element-wise multiplication between matrices. The approximate solution of $\phi$ at $t+\Delta t$ can be obtained by taking inverse Fourier transform:
\[
\Phi(t+\Delta t) = \textbf{iFFT}(\hat{\Phi}(t+\Delta t)).
\]
Similarly for the \eq \ref{eq:rho}, we can write it into the semi-implicit form in terms of $\phi\rho$,
\begin{align*}
(\phi\rho)(\vb{x}, t&+\Delta t)-\text{Pe}^{-1}\Delta t \nabla^2(\phi\rho)(\vb{x},t+\Delta t) =\\
& (\phi\rho)(\vb{x},t) -\text{Pe}^{-1}\Delta t \Big(\nabla\cdot(\rho\nabla\phi)\Big)(\vb{x},t) + (\phi f)(\vb{x},t)
\end{align*}
and then apply the \textbf{FFT} and \textbf{iFFT} to find the approximate solution of $\phi\rho$ at $t+\Delta t$. The approximate solution $U = (\rho_{ij})$ at $t+\Delta t$ is obtained by:
\begin{align*}
\rho_{ij} = 
\begin{cases}
(\phi\rho)_{ij}/\phi_{ij}, & \text{if}\ \phi_{ij}\ge 10^{-4},\\
(\phi\rho)_{ij},  & \text{if}\ \phi_{ij}<10^{-4}.
\end{cases}
\end{align*}

In our numerical simulations, we take $L_x = L_y = 2.5$, $N_x = N_y = 256$ and $\Delta t = 5\times 10^{-4}$.

\subsection{Initial Conditions}
Initially the cell shape is taken to be circular with radius $R_0 = 1$, $\phi = \frac{1}{2} + \frac{1}{2}\tanh \left[3(R_0-r)/\epsilon\right]$ with $\epsilon = 0.1$, and the $\rho$-distribution equal to 0.8 in the front half and 0 in the rear half, with an added normally-distributed noise with standard deviation of 0.2.  

\section{Sharp interface models without the quasi-circularity assumption}

\subsection{Exact Calculation of Steady-state Shapes of Straight Cells}\label{subsection:exact_calculation}

\begin{figure}[ht!]
\includegraphics[height=50mm]{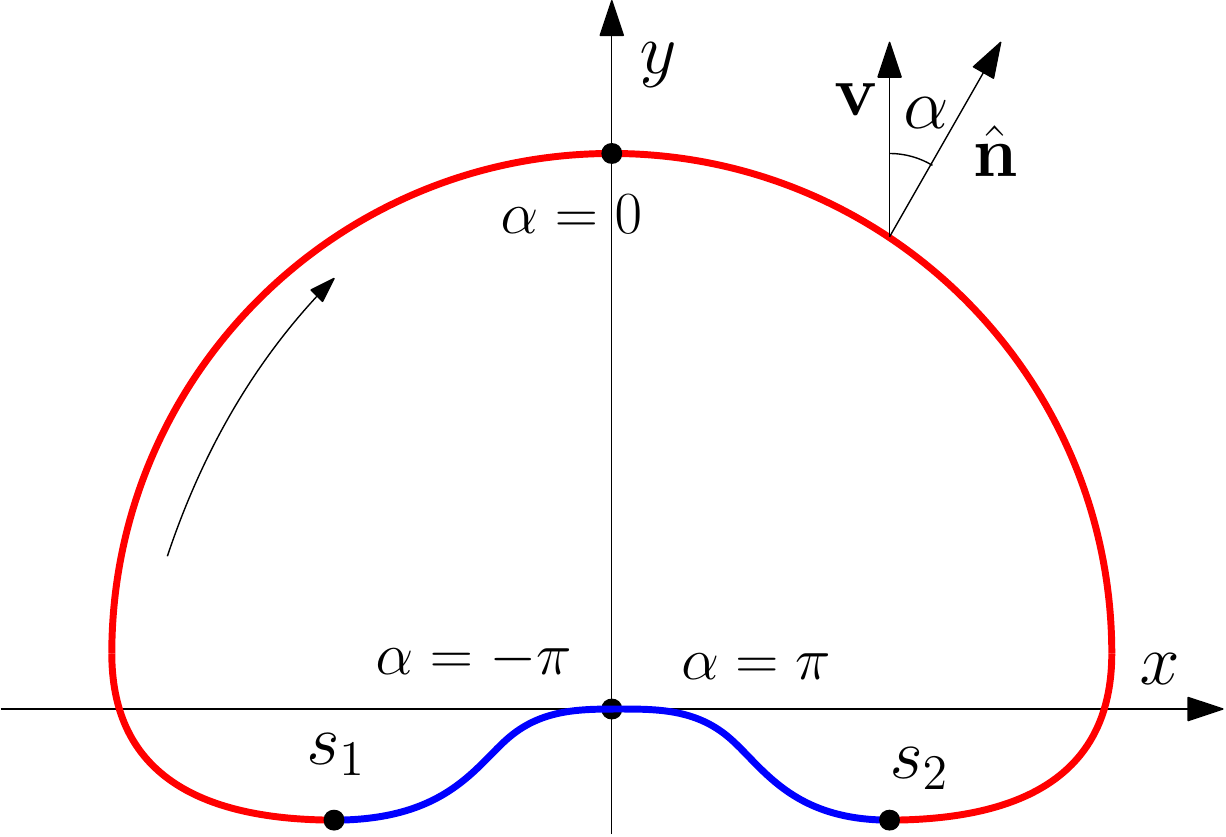}
\caption{Schematic for the steady state of a straight crawling cell in the sharp interface model. The curved arrow inside indicates the increasing of arc-length $s$. The ratio of the width and height of the circumscribed rectangle defines the aspect ratio of the cell. }  \label{fig:straight_sharp_schematic}
\end{figure}

We can find a numerically exact solution to the steady-state shape of the cell by integrating the sharp-interface equation \eq \ref{eq:sharp_interface_limit_normal_velocity}; related approaches, including graded radial extension models, have been applied before \cite{keren2008mechanism,adler2013closing}. If we assume that the cell is crawling along a straight trajectory with constant velocity $\textbf{v} = v \hat{\textbf{y}}$, where $\hat{\textbf{y}}$ stands for the unit $y$-axis direction, 
then the relation between the cell velocity and cell shape can be obtained from \eq \ref{eq:sharp_interface_limit_normal_velocity}:
\begin{align}\label{eq:straight_sharp}
v\cos\alpha = (\rho - \rhoz) - \chi\dot{\alpha}
\end{align}
where $\alpha$ is the angle counterclockwise from  $\hat{\textbf{n}}$ to $\textbf{v}$, see \fig \ref{fig:straight_sharp_schematic}, and the overhead dot represents the derivative with respect to the arc-length $s$. We take the boundary conditions
\begin{align}
\alpha(0) = -\pi, \alpha(L) = \pi \label{eq:straight_sharp_bc}
\end{align}
which are appropriate if the cell perimeter is a simple closed curve, and there is no cusp at $s = 0$ (\fig \ref{fig:straight_sharp_schematic}).  Given a solution $\alpha(s)$, we can find the cell shape by integrating the tangent vector along the arclength, 
\begin{equation}
\dot{x} = \cos\alpha,\quad \dot{y} = -\sin\alpha.
\end{equation}
Because the cell is a simple closed curve, we will find that $x(0) = x(L)$ and $y(0) = y(L)$. 

The only unknown parameter in \eq \ref{eq:straight_sharp} is the cell's velocity $v$, if we specify the cell perimeter.  The two boundary conditions of \eq \ref{eq:straight_sharp_bc} on the first-order equation of \eq \ref{eq:straight_sharp} shows that, for a fixed value $L$, there will be only a particular value of $v$ for which \eq \ref{eq:straight_sharp} can be solved.

\begin{figure*}[ht!]
\includegraphics[width=120mm]{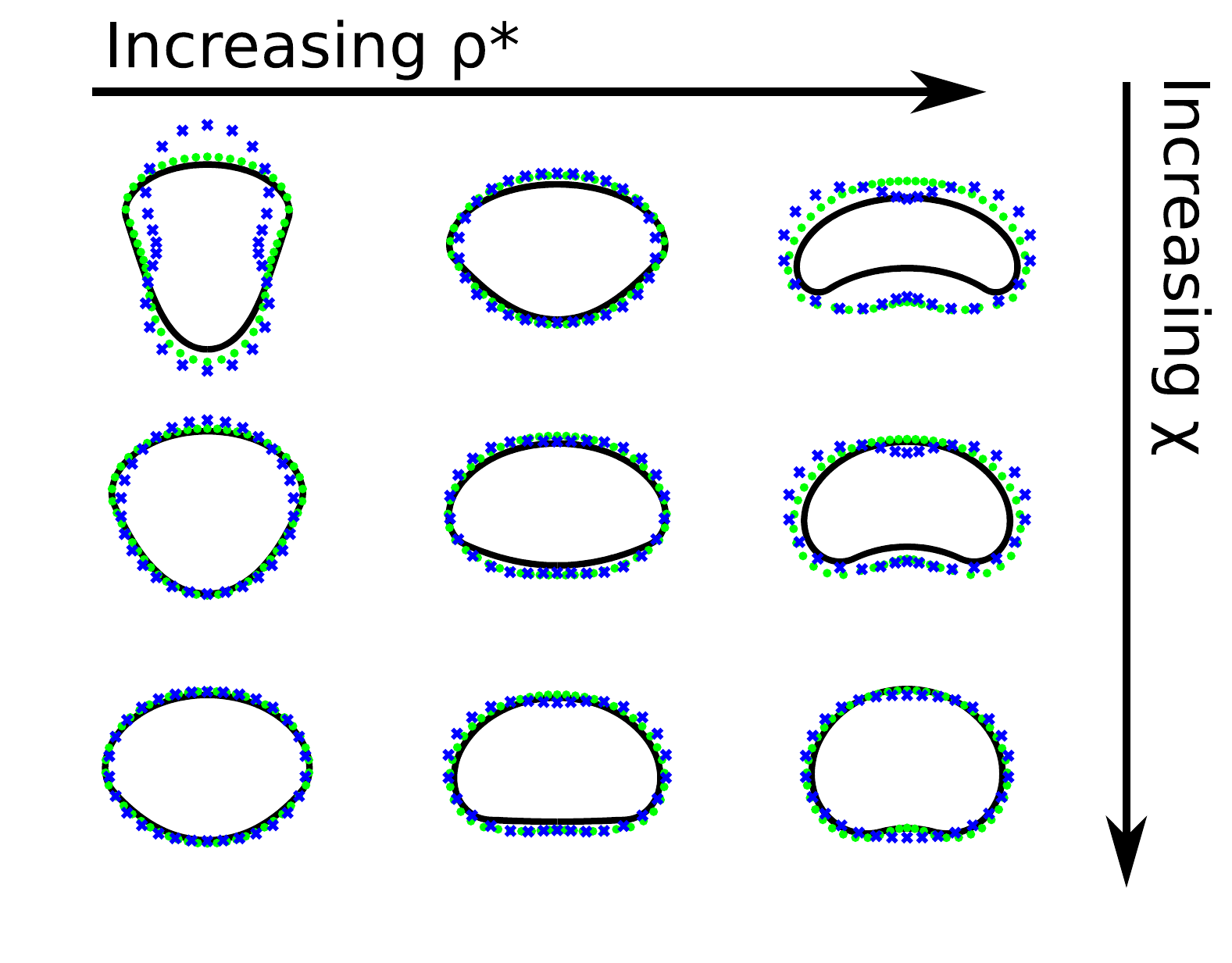}
\caption{Comparison between numerically exact calculation of sharp-interface cell shapes and the quasicircular approximations. Solid lines are the numerically exact calculation, blue crosses are the quasicircular model with $N = 2$, and green dots are the quasicircular model with $N = 100$.  In this figure, cells are shown with their velocity along the positive $y$ direction.  The values of $\chi$ are $\chi = 0.2,0.3,0.4$ and $\rhoz = 0.2, 0.35, 0.5$.  $L = 2\pi$ is the perimeter for the numerically exact results, and $R_0 = 1$ for the quasicircular calculations.  Details of numerical calculation are in Section \ref{subsection:exact_calculation}.}  \label{fig:Fourier_sample}
\end{figure*}

We know from our simulations and the analysis of \cite{mori2008wave, mori2011asymptotic} that $\rho$ has a sharp interface between values $\rho^+$ and $\rho^-$:
\begin{align}
\rho(s) = 
\begin{cases}
\rho^+,\quad & s_1 < s <  s_2  \\
\rho^-, \quad &  s\le s_1 \ \text{or}\ s\ge s_2
\end{cases}
\end{align}
where $s_1 = (L-L^+)/2$, $s_2= (L+L^+)/2$ and $L^+$ is the length of the $\rho^+$ region, see  \fig \ref{fig:straight_sharp_schematic} in which the red part stands for the $\rho^+$ region. How large is $L^+$? Integrating \eq \ref{eq:straight_sharp} along the arc-length and using \eq \ref{average_vn_zero}, we find that
\begin{align*}
2\pi\chi = \int_0^L ds (\rho-\rhoz) = L^+\rho^+ + (L-L^+)\rho^- - L\rhoz,
\end{align*}
and then
\begin{align}\label{eq:L1}
L^+ = \dfrac{2\pi\chi -L(\rho^- - \rhoz)}{\rho^+-\rho^-}.
\end{align}
In practice, $\rho^+ = 1$ and $\rho^- = 0$ are appropriate for our reaction-diffusion equations.

\eq \ref{eq:straight_sharp} coupled with the boundary conditions of \ref{eq:straight_sharp_bc} is numerically solved by using a shooting method to determine the value of $v$ for which the equations can be solved.  As an example, we choose $L = 2\pi$.  We compare this result with our quasicircular approximation in \fig \ref{fig:Fourier_sample}, seeing generally good agreement on cell shape and size.  Even using the model with only $N = 2$ Fourier modes, we capture the appropriate trends of cell shape.  We choose $R_0 = 1$ in the quasicircular approximation (\eq \ref{Fourier_expansion}); this is set so that the cell's contour length $L$ in the quasicircular approximation $L \approx 2\pi R_0$ matches that for the numerically exact method. (We note that the agreement between contour lengths is only approximate, and the contour length of some cells shown in \fig \ref{fig:Fourier_sample} can deviate from $2\pi$.)  Velocities predicted by the Fourier model are also in good agreement with those predicted by the numerically exact model, with errors of less than 25\% for the parameters shown in \fig \ref{fig:Fourier_sample}, even only including $N = 2$ modes.  

\begin{figure*}[ht!]
\includegraphics[width=140mm]{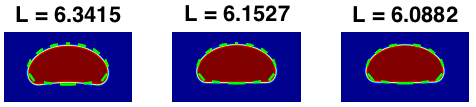}
\caption{Comparison between sharp interface solution and phase field solutions. The parameter $K=500,2000,4000$ from left to right. The other parameter values are $\chi = 0.26$, $\text{Pe}^{-1} = 0.30$, $\rhoz = 0.40$, $C = 6$, $m = 0.5$ and $\epsilon = 0.1$.   Color plots indicate the phase field, dashed line the sharp-interface results.  Lengths for each value of $K$ (listed above the image) are taken from the corresponding phase field simulations.}  \label{sharp_vs_diffuse}
\end{figure*}

The sharp interface solutions also have good agreement with the phase field solutions (\fig \ref{sharp_vs_diffuse}).  However, this agreement depends on the validity of our assumptions, e.g. that the value of $\rho$ at the cell front is unity.  This is controlled by $K$.  Mathematically, $K$ resembles a penalty constant to maintain $\rho^- = 0, \rho^+ = 1$ (\eq \ref{eq:rho}). As $K$ becomes greater, $\rho^+$ becomes closer to $1$ and the phase field cell shrinks to the sharp interface one.

\subsection{Exact Calculation of Steady-state Shape of Cells with Circular Trajectories}

\begin{figure}[ht!]
\includegraphics[height=45mm]{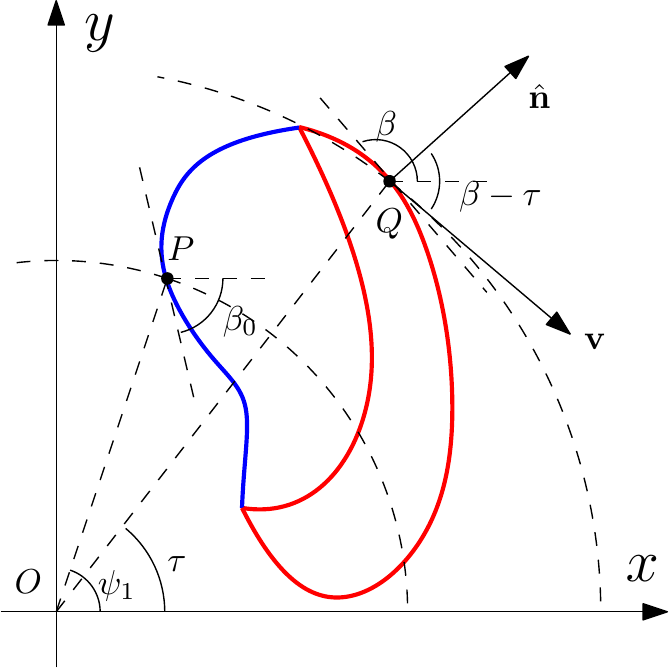}
\caption{Schematic for the steady state of a circularly crawling cell in the sharp interface model. }  \label{circular_sharp_schematic}
\end{figure}

\begin{figure}[ht!]
\includegraphics[height=60mm]{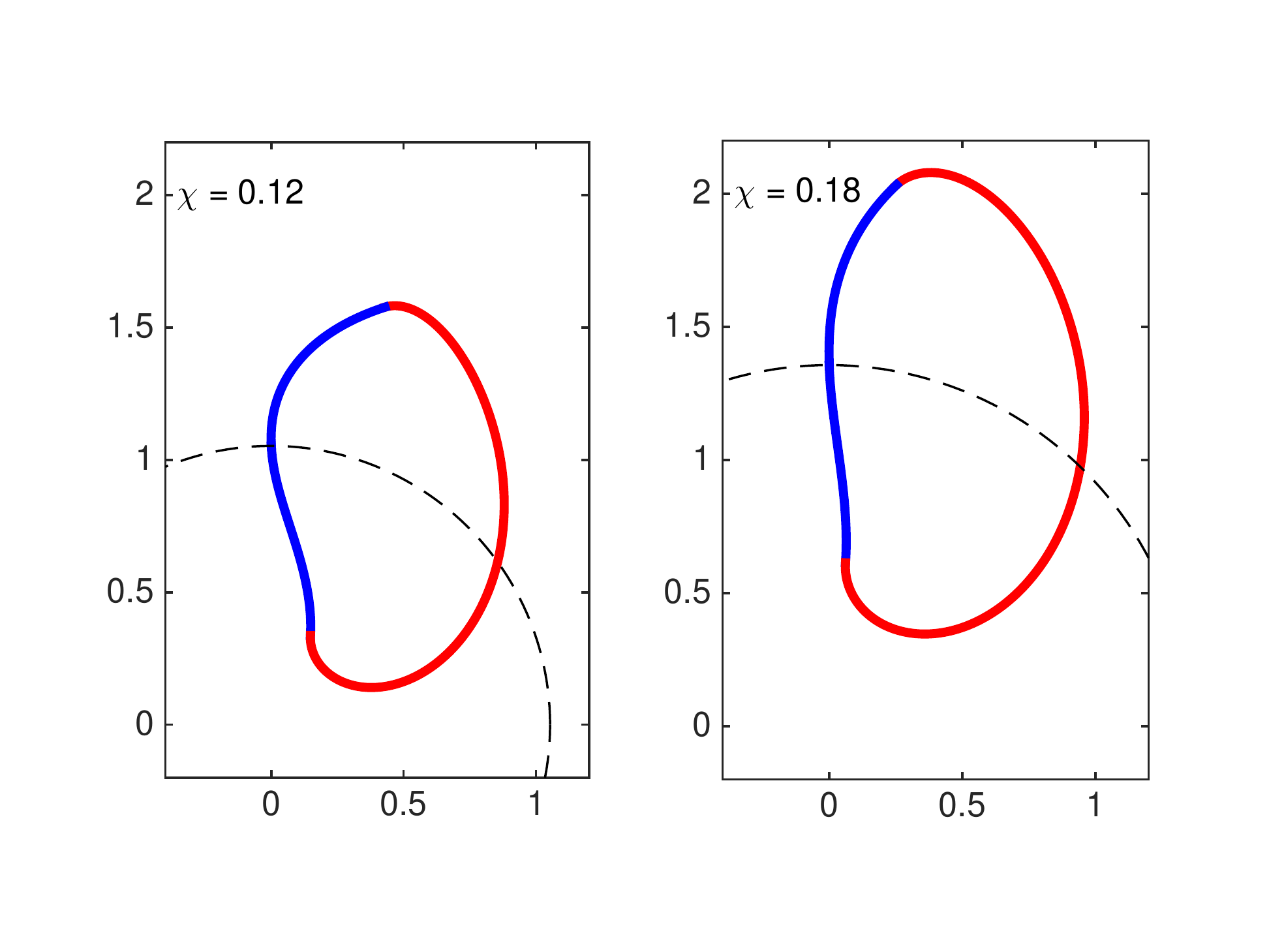}
\vspace{-0.4in}
\caption{(Color online) Exactly calculated sharp-interface cell shape in circular steady state. The tension $\chi = 0.12$ (left) and $0.18$ (right). The values of angular velocity $\omega_0$ and the radius of the circular orbit are obtained from phase field simulations (\ref{eq:pfu}-\ref{eq:rhocyt}): $\omega_0 = 0.5775$ (left) and 0.3832 (right), $R_0 = 1.0530$ (left) and 1.3563 (right). Other parameters are $\rhoz = 0.4$.}  \label{circular_sharp_sample}
\end{figure}

While we only used the sharp-interface results in the main paper to determine the steady-state shape of cells on straight trajectories, our techniques can also be used to determine the shape of cells undergoing circular motion.
Schematically, we show a circularly moving cell in steady state in  \fig \ref{circular_sharp_schematic}. The origin $O$ is the center of rotation, $Q$ is a representative point on the cell boundary, $\tau$ is the angle made by $OQ$ from horizontal, $\beta$ is the angle of inclination from horizontal, $\hat{\textbf{n}}$ is the normal to the cell boundary, and $\textbf{v}$ is the moving direction of $Q$ which is normal to $OQ$. The blue curve represents the $\rho^-$ region and the red part is $\rho^+$ region.  The point $P$ is the middle point of the blue $\rho^-$ region with respect to arc-length and is taken as the starting point for arc-length parameterization, namely, $s = 0$ at $P$.  Note that the cell shape is independent of angle $\psi_1$, so in the sharp-interface model, we take $\psi_1 = \pi/2$ to keep $P$ on the $y$-axis.  The angle between $\hat{\textbf{n}}$ and $\textbf{v}$ is $\beta-\tau$ and the normal velocity
\begin{align}
v_n = \omega_0\sqrt{x^2 + y^2}\cos(\beta-\tau),
\end{align}
where $\omega_0$ is the constant angular velocity for the cell in circular steady state.
Then in the circular steady state, we have
\begin{align}
&\ \omega_0\sqrt{x^2+y^2}\cos(\beta-\tau) = (\rho-\rhoz) - \chi\dot{\beta}\nonumber\\
\Rightarrow&\ \omega_0\sqrt{x^2+y^2}\ (\cos\tau\cos\beta+\sin\tau\sin\beta) = \rho-\rhoz - \chi\dot{\beta}\nonumber\\
\Rightarrow&\ \omega_0\ (x\cos\beta+y\sin\beta) = \rho-\rhoz - \chi\dot{\beta}.\label{Eqn_theta}
\end{align}
Notice that 
\[
x\cos\beta + y\sin\beta = \dfrac{1}{2}(x^2+y^2)^{\cdot},
\]
Integrating Eq. \ref{Eqn_theta} along the arclength, we find that
\[
2\pi\chi = \int_0^L (\rho-\rhoz)\ \text{d}s,
\]
exactly as what we obtain from the equation for the straightly moving cell.

Finally we obtain the system for steady state of circular cells:
\begin{align}\label{eq:circular_sharp}
&\omega_0\ (x\cos\beta+y\sin\beta) = \rho-\rhoz - \chi\dot{\beta},\\
&\dot{x} = \cos\beta,\quad \dot{y} = \sin\beta, \nonumber
\end{align}
with the following boundary conditions
\begin{align}
\beta(0) &= \beta_0, &  x(0) &= 0, & y(0) &= R_0. \label{eq:circular_sharp_bc1}\\
\beta(L) &= \beta_0 + 2\pi, & x(L) &= x(0), & y(L) &= y(0). \label{eq:circular_sharp_bc2}
\end{align}
Note that for the boundary conditions (\ref{eq:circular_sharp_bc2}) at $s = L$, $x(L) = x(0)$ and $y(L) = y(0)$ automatically implies $\beta(L) = \beta_0 + 2\pi$ by considering (\ref{eq:sharp_interface_limit_normal_velocity}), (\ref{average_vn_zero}) and (\ref{eq:circular_sharp}), so $\beta(L)$ is a redundant boundary condition. Besides, the cell perimeter $L$ and the initial inclination angle $\beta_0$ are two unknown parameters, so the number of unknowns matches with the number of boundary conditions in (\ref{eq:circular_sharp}-\ref{eq:circular_sharp_bc2}). The length $L^+$ of the $\rho^+$ region is determined by \eq \ref{eq:L1}, which is treated known once $L$ is determined.

Numerically, we solve the \eq \ref{eq:circular_sharp} coupled with boundary conditions (\ref{eq:circular_sharp_bc1}-\ref{eq:circular_sharp_bc2}) by a shooting method. The parameter values are $\rhoz = 0.4$, $\chi$ = 0.12 (0.18, respectively), $R_0$ = 1.0530 (1.3563, respectively) and $\omega_0$ = 0.5775 (0.3832, respectively) where the cell's angular velocity $\omega_0$ and circular radius $R_0$ are taken from the phase field simulations  (\ref{eq:pfu}-\ref{eq:rhocyt}). As in the earlier sharp interface limit calculations, we take $\rho^+ = 1, \rho^- = 0$. The numerical results are presented in \fig \ref{circular_sharp_sample} which have good agreement with the phase field simulations.


\end{document}